%% file: main.tex
\documentclass[conference]{IEEEtran}
\usepackage{amsmath,amssymb,amsfonts}

\usepackage{framed,subfigure}
\usepackage{graphicx}
\usepackage{textcomp}
\usepackage{multirow}
\usepackage{xcolor}
\usepackage{makecell}
\usepackage{hyperref}
\usepackage{paralist}
\usepackage{bbding}
\usepackage{fullpage}
\usepackage{times}
\usepackage{fancyhdr,graphicx,amsmath}
\usepackage[ruled,vlined,linesnumbered]{algorithm2e}
\usepackage{bm}
\usepackage{float}

\usepackage{amsmath}
\usepackage{xcolor}
\usepackage{colortbl}
\usepackage{multirow}
\usepackage{subfigure}
\usepackage{todonotes}
\usepackage[noend]{algpseudocode}
\usepackage{makecell}
\usepackage{booktabs}

\algrenewcommand\algorithmicrequire{\textbf{Input:}}
\algrenewcommand\algorithmicensure{\textbf{Output:}}

\renewenvironment{thebibliography}[1]{%
\begin{oldthebibliography}{#1}%
    \setlength{\parskip}{0\baselineskip}
    \setlength{\itemsep}{0\baselineskip}
}%
{%
\end{oldthebibliography}%
}
\begin{document}

\title{QCFE: An efficient Feature engineering for query cost estimation.} 

\author{\IEEEauthorblockN{Yu Yan,Hongzhi Wang, Junfang Huang, Dake Zhong, Man Yang, Kaixin Zhang, Tao Yu, Tianqing Wang}
\IEEEauthorblockA{\textit{Harbin Institute of Technology Harbin China, HUAWEI China } \\
yuyan@hit.edu.cn,wangzh@hit.edu.cn\\} }


\maketitle

\begin{abstract}

   Query cost estimation is a classical task for database management. Recently, researchers apply the AI-driven model to implement query cost estimation for achieving high accuracy. However, two defects of feature design lead to poor cost estimation accuracy-time efficiency. On the one hand, existing works only encode the query plan and data statistics while ignoring some other important variables, like storage structure, hardware, database knobs, etc. These variables also have significant impact on the query cost. On the other hand, due to the straightforward encoding design, existing works suffer heavy representation learning burden on ineffective dimensions of input. To meet the above two problems, we first propose an efficient feature engineering for query cost estimation, called QCFE. Specifically, we design a novel feature called feature snapshot to efficiently integrate the influences of the ignored variables. Further, we propose a difference-propagation feature reduction method for query cost estimation to filter the useless features. The experimental results demonstrate our QCFE could largely improve the time-accuracy efficiency on extensive benchmarks.
\end{abstract}

\section{Introduction}
\label{introduction}
    \input{content/introduction}

\section{overview}
\label{sec:overview}
\input{content/overview}

\section{The snapshot feature}\label{sec:model}

\input{content/know}

\section{Feature Reduction}\label{sec:featureselection}
\input{content/selection}

\section{The Evaluation Of QCFE} \label{sec:evaeerl}
\input{content/experiment_results}

\section{Related Works}
    \label{related}
        \input{content/related}

\section{Conclusion}
\label{conclusion}
    \input{content/conclusion}

\newpage
\bibliographystyle{ieeetr}
\bibliography{citations/bibliography}

\end{document}

%% file: content/introduction.tex
Cost estimation plays a pivotal role in database management, forming the bedrock for database optimization strategies encompassing query optimization~\cite{zhou2021sia}, index optimization~\cite{reindex}, and storage efficiency, among other aspects. The precision of cost estimation methodologies stands as a linchpin for achieving optimal performance in database operations. Regrettably, conventional techniques reliant on cost equations may produce huge estimation errors under complex workloads due to their simplistic frameworks and underlying assumptions~\cite{2022zero}. This shortcoming can engender sub-optimal optimization outcomes, consequently undermining database performance.

Hence, the database community has delved into the application of neural network models to capture the intricate correlation between queries and their associated costs, harnessing the formidable learning prowess inherent in these deep networks. Extensive experimental results~\cite{card1,card2, card3, 2022zero} have demonstrated that the learning approaches achieve high accuracy across various complex benchmarks.

However, the utilization of neural networks for databases poses an efficiency-accuracy dilemma. On one hand, the query cost is related to multiple features, such as relation table, query, etc.), compared to other fields, such as natural language processing~\cite{chowdhary2020natural} and image recognition~\cite{wu2017new, Vu_2018_ECCV_Workshops}, as they contain different structure. To accurately represent and fit the query plan-performance relationship, it is necessary to design complex network models. For instance, the transformer query cost model~\cite{queryformer} has shown superior performance compared to simpler models, primarily due to its deeper network layers and attention mechanism that assigns weights to features. Notably, such network structures require more training and inference time than ordinary deep neural network (DNN) models. 

On the other hand, database cost estimation is a frequently invoked component, which may be invoked multiple times within a single query optimization~\cite{ioannidis1996query}. For example, even PostgreSQL utilizes genetic algorithms to reduce the number of estimation requests, the cost estimation component is still called more than $n$ times, where $n$ is the number of nodes in the query plan. Additionally, large database management systems receive thousands of query requests every minute. Consequently, it is not feasible to allocate excessive time to the fundamental cost estimation component.

\begin{figure}[htb]
    \centering
    \includegraphics[width=0.9\linewidth]{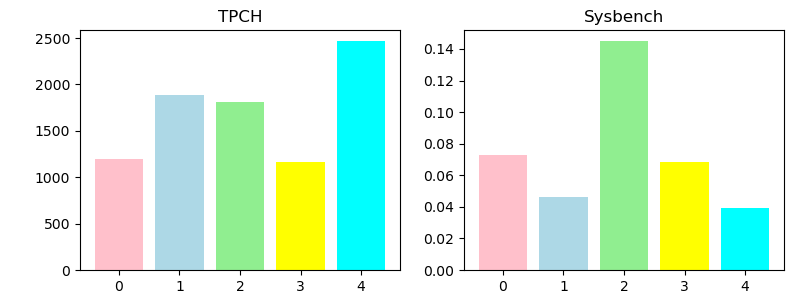}
    \caption{The average query cost (ms) of 1000 queries in TPCH and Sysbench under different database environments.}
    \label{fig:execution}
\end{figure}

Given the limitations of designing complex models, for solving the efficiency-accuracy dilemma, a natural approach should optimize the features to reduce the burden of representation learning and fitting learning in the model. Existing AI-driven cost estimation methods utilize relatively straightforward approaches to processing input query features. Typically, the one-hot encoding for tables~\cite{sun2019end}, the one-hot encoding for indexes~\cite{marcus2019plan}, and the vector for numerical values are directly fed into the evaluation model in a bottom-up path of the query. Totally, we have identified two shortcomings regarding feature design in existing AI-driven query cost estimators:

(1) \textbf{Missing Important Features:} Current methods~\cite{2022zero} primarily focus on encoding the query plan and the table statistics, often overlooking the impact of other database variables on query cost. However, variables such as the storage format of data (e.g., B+ tree or LSM tree) and the hardware of the database also play a significant role in determining the query cost. Our investigation, as depicted in Figure~\ref{fig:execution}, demonstrates substantial differences (2 times in TPCH and 3 times in Sysbench) in the average execution time of the same queries under different database environments (five database knob configurations). Therefore, neglecting the database environment can result in significant losses when predicting query cost.

(2) \textbf{Heavy Representation Learning:} Existing methods directly utilize the table feature, index feature, operator feature, etc. as the input of the AI cost model. This brings a large burden for representation learning~\cite{zhang2018network}, which is used to learn the effective representation of input features.
Specifically, with the goal of simplifying learned model (accelerating inference time), capturing the relationships between the large amount of features and the query cost can be a difficult task. This intricate logic relationship between multiple features necessitates multiple nonlinear transformations to effectively capture.

These two questions appear to be a contradiction. The absence of crucial features primarily results from an incomplete modeling of the query cost estimation problem, necessitating the incorporation of additional features. The heavy representation learning stems from the ineffective elements of the encoding, necessitating the removal of some features. Nevertheless, when viewed collectively, these issues can be categorized as feature engineering challenges, implying that the task of query cost estimation's feature engineering has not been processed optimally.

To solve the above problems, we design an \emph{effective \underline{f}eature \underline{e}ngineering for \underline{q}uery \underline{c}ost estimation}, called QCFE. The core sights are as follows:
(1) To avoid missing important variables, we define a novel concept, called \textbf{feature snapshot ($SF$)} to integrate the characteristics of ignored variables (defined as the variable set of database knobs, storage structure, hardware and operating system). To the best of our knowledge, no one has attempted to encode the ignored variables for the query cost model. One possible reason may be that the
resource required to build an exact feature representation is tantamount to build the database environment. Hence, we propose an estimated method to obtain the snapshot feature, ensuring high efficiency.

(2) For the heavy representation learning, we design a difference-propagation feature reduction (FR) method, to relieve the learning burden by pruning the ineffective features. Specifically, depending on the relational table and load type, certain features may not be effective. For instance, the plan method employs columns with the attribute's length to encode the index. However, in pure write scenarios, the database management system may not create an index, resulting in an ineffective feature with the length of the number of columns in the query feature. These ineffective features not only increase the training and inference cost of the AI evaluation model but also reduce its accuracy~\cite{dong2018feature}.

Totally, the specific contributions are as follows:

\begin{itemize}
    \item In order to improve the time-accuracy efficiency, we propose a feature engineering for query cost estimation, called QCFE. 
    \item We first propose the feature snapshot (in Section~\ref{sec:model}) concept for query cost estimation, integrating  the influential the ignored variables variables. Our core goal is to make some reasonable assumptions to calculate the feature snapshot with high time efficiency. 
    
    \item  We design the difference-propagation feature reduction method (in Section~\ref{sec:featureselection}) to efficiently reduce the useless feature, further improving model training and inference efficiency.  
    \item To clarify the effectiveness of our QCFE, we demonstrate various comparisons (Section~\ref{sec:evaeerl}) under extensive popular benchmarks (TPC-H, job-light, and Sysbench), including the evaluation of time-accuracy efficiency, the ablation of QCFE, the robustness of QCFE, etc.
\end{itemize}

%% file: content/overview.tex
In this section, we overview the architecture and workflow of our QCFE.
\begin{figure}[htb]
    \centering
    \subfigure[The overview of QCFE.]{
		\includegraphics[width=0.6\linewidth]{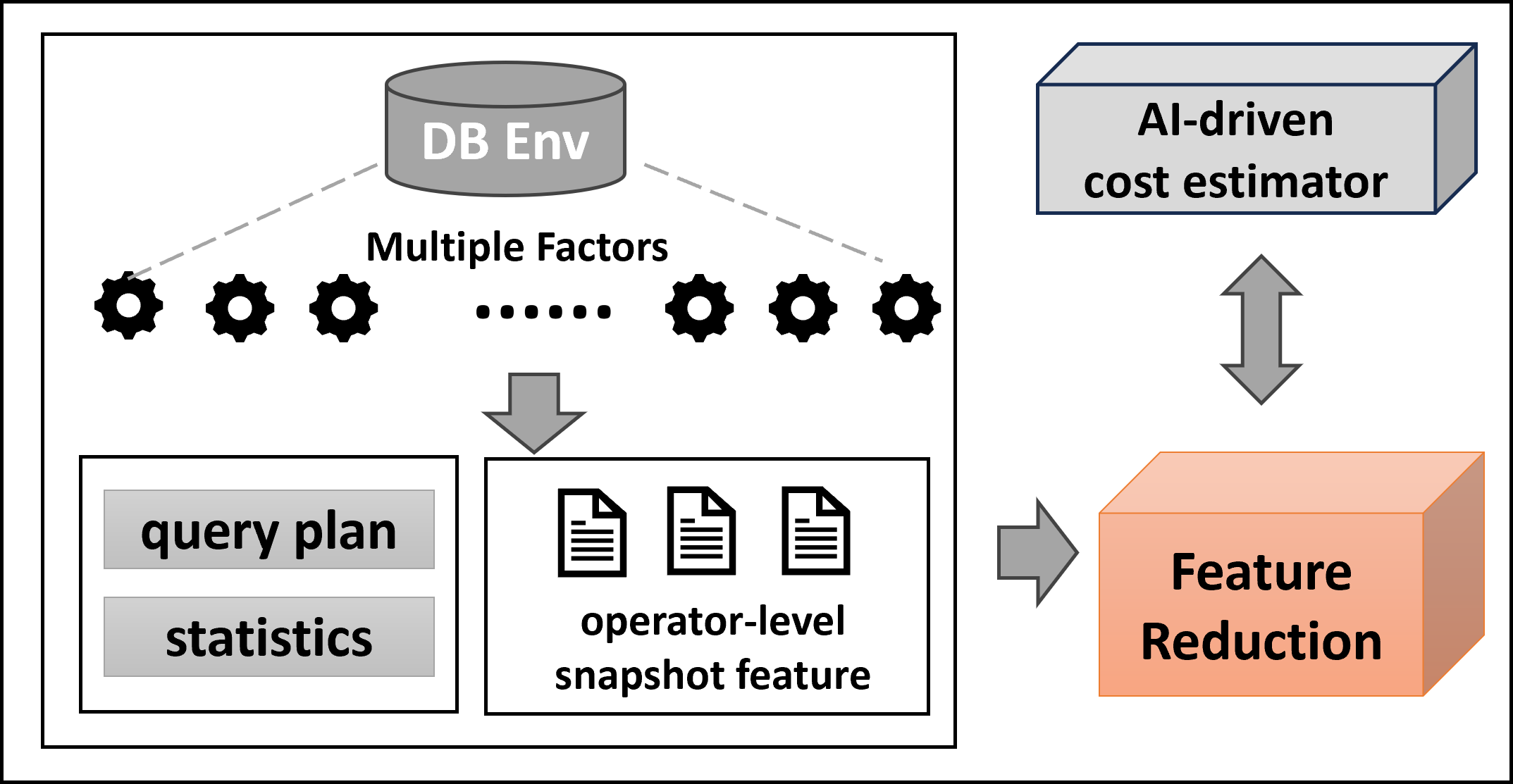}
		\label{fig:qcfe}
	}
    \subfigure[General FE.]{
		\includegraphics[width=0.32\linewidth]{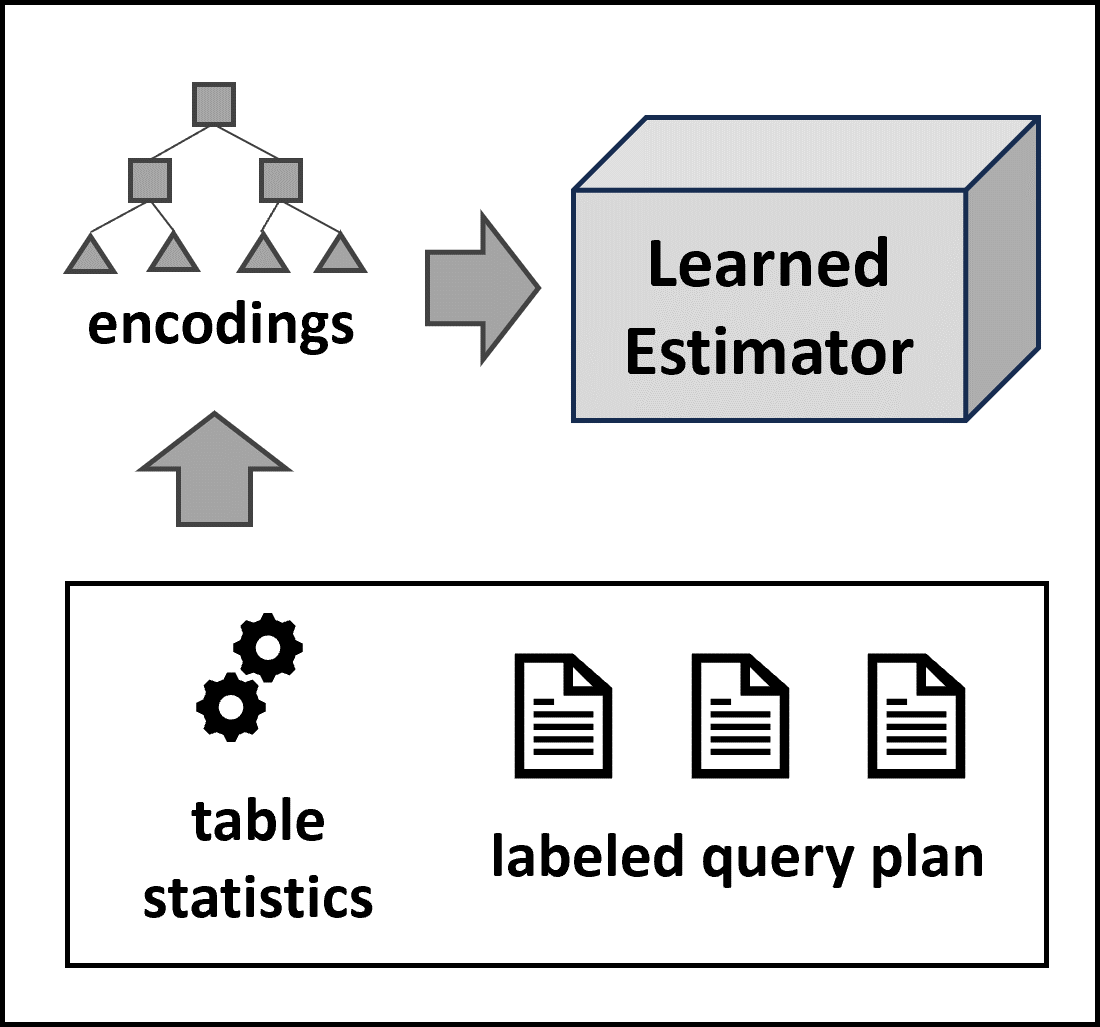}
		\label{fig:ofe}
	}

    \caption{QCFE VS General feature engineering.}
    \label{fig:overview}
\end{figure}

Firstly, we show the general feature engineering which is widely used in existing works~\cite{view,sun2019end,mscn}, to clarify the effectiveness of our QCFE. 
As shown in Figure~\ref{fig:overview}, the general FE directly encodes the query plan and table statistics as the training set for AI-driven models, which ignores some other Influential factors (like hardware) and may have meaningless computing resource overhead of ineffective input codes.

In contrast, our QCFE considers all the factors of the database environment containing the table statistics, queries, hardware, database knobs, etc. Especially, we design an efficient feature snapshot to capture the ignored variables influence in Section~\ref{sec:model}. Then, we propose the difference propagation feature reduction algorithm (a variant of the back-propagation algorithm~\cite{back}) to filter the useless dimensions to further improve time-accuracy efficiency in Section~\ref{sec:featureselection}.

%% file: content/know.tex
Although the ignored variables also have significant impacts on the query cost, it involves resource overhead to make exact representations for these variables as a feature snapshot. In this section, we introduce the sophisticated feature snapshot design in Section~\ref{sec:analysis}. Further, to improve the time efficiency of calculating the feature snapshot, we design a standard simplified SQL template to replace the original templates.

\subsection{Estimated Feature Snapshot}\label{sec:analysis}

Firstly, we clarify why we design an estimated FS to represent the ignored variables instead of an exact feature from the following two reasons.

(1) \textbf{Partial influence:} Essentially, these ignored variables affect the query cost by affecting the I/O cost and CPU cost of the operators. Only partial components of these variables have impact on the query cost. For example, The audio peripherals in the hardware will not affect the query execution efficiency. Hence, we only need to consider the I/O and CPU-related partial components. 

(2) \textbf{High resource overhead:} The ignored variables have complex structures, which are costly to be directly encoded as an exact feature. For example, the hardware consists of multiple components, such memory, disk, CPU, etc. The exact representation of the various components of the ignored variable will bring enormous feature snapshots, leading to large space cost.

Due to the partial influence and high resource overhead, we consider to construct an estimated feature snapshot representation for the ignored variables. The specific considerations are as follows: 

(1) We firstly identify how the ignored variables affect the query cost. Specifically, we analyze a basic physical cost formula of PostgreSQL, $Cost_{total} = c_s $ (the I/O to sequentially access a page) $ \times n_s + c_r $ (the I/O cost to randomly access a page) $ \times n_r + c_t $ (the CPU cost to process a tuple) $\times n_t + c_i$ (the CPU cost to process a tuple via indexes)$ \times n_i +c_o $ (the CPU cost to process an operator, like sort)$\times n_o$. We can conclude two metrics from the above formula, the cost coefficient ($C = \{c_s, c_r,c_t,c_i,c_o\}$) and the cost number($N = \{n_s,n_r,n_t,n_i,n_o\}$). All the ignored variables  influence the query cost by influencing the $C$ and $N$. In general, the query plan and data statistics have the main impact on $N$ while the ignored variables mainly influence the cost of once I/O and CPU of request $C$. For example, the disk type only influences the I/O speed for a given relation and query. And the enable index scan knob mainly influences the $C$ in once request. Our first estimation is that the ignored variables only influence the $C$.

\begin{table}[h!]	
    \renewcommand{\arraystretch}{1.1}
        \centering 
        \caption{The logical formula knowledge.}  
        \label{tab:know} 
    \begin{tabular}{|m{2.5cm}|m{4cm}|}
        \hline
        Cost Formula & Operators \\
        \hline
        $F=c0\times n+c1$ & Seq Scan, Materialize, Aggregation \\
        \hline
        $F=c0\times n+c1$ & Index Scan, Merge Join, Hash Join\\
        \hline
        $ F=c0\times nlogn+c2$ & Sort \\
        \hline
        $F=c0\times n1\times n2+c1\times n1+c\times 2\times n2+c3$ & Nested Loop \\
        \hline
    \end{tabular}
    \end{table}
 
(2) We present how to estimate the $C$. For simplicity and generality, we leverage the notion of logical cost functions~\cite{logical} to estimate the $C$ instead of the cost formula of certain DBMS. Table~\ref{tab:know} shows the specific logical assumptions of the cost formula for different operators. Here, these formulas is determined by the logical execution of operators. For example, $F=c0 \times n+c1$ could be the logical formula for seq scan~\cite{2014Uncertainty}, where $n$ is the cardinality of operator. Based on these assumed logical formula, we could estimate the $C$ for each operator instead of an incredible exact representation, representing the influences of the ignored variables. For example, $[c0,c1]$ could be the feature snapshot of the seq scan operator. Also, these assumed logical formulas could be optimized for specific DBMS (such as the revised cost formula for Postgres~\cite{improve}), improving FS estimation accuracy.

 Specifically, according to the logic formulas, we utilize the regression model, the least square method~\cite{bjorck1990least} to calculate the FS for each operator. The regression model is trained from the labeled operator set, which is collected by multiple query executions. When the query structure is complex, collecting labeled sets may be costly in time.

\subsection{Simplified Template}\label{sec:assume}

We observe that calculating the feature snapshot needs to execute multiple queries, which is costly for  complex queries such as the OLAP in TPCH. In order to improve time efficiency, we design simplified templates, which could not only capture the characteristic but also have efficient execution time.

\begin{algorithm}
    \SetKwInOut{Input}{input}
    \Input{
        \ the data abstract ($R$), the original query templates ($P$) and the scale of transferred templates ($N$)\\
    }
    \SetKwInOut{Output}{output}
    \Output{
        \ $Q$ is an effective substitute of the original queries. \\
    }
    \BlankLine

    $info, T, Q \leftarrow \varnothing, \varnothing, \varnothing $\\

    \For(){$p \in P$}{
        \For(){$s \in p.split()$}{
            \If{$s.match()$}{
                $info[s.op].add([s.tab, s.col])$ // \textit{update the operator-table-column information}\\
            }
         }
    }
    \For(){$op \in info$}{
      
        \For() {$(t,c) \in info[op]$}{
            $T_{op} \leftarrow genTemplate(op,t,c)$ // \textit{generate template according to op}\\
        $ T \leftarrow T \cup T_{op}$ \\
        }
    }
    \While{$N > 0$}{
    \For(){$t \in T$}{

        $ q \leftarrow setRandomValue(R, t, random(<,>,=,...))$ // \textit{fill template with R and a random operator keyword}\\
        $ Q \leftarrow Q \cup q$ \\
    }
    $N \leftarrow N -1$
    }
    
    return $Q$ \\

\caption{Generate the simplified query templates.}\label{alg:know}

\end{algorithm}

As shown in Algorithm~\ref{alg:know}, the input consists of the data abstract $R$, the original query templates $P$, the scale of transferred template $N$, and the output is a query set as an effective substitute for calculating FS. Our algorithm consists of three important phases, (i) Parse original Query Templates. (ii) Generate simplified templates. (iii) Fill in simplified templates. Next, we introduce these phases in detail.
\begin{figure*}[htb]
    \centering
    \includegraphics[width=0.9\linewidth]{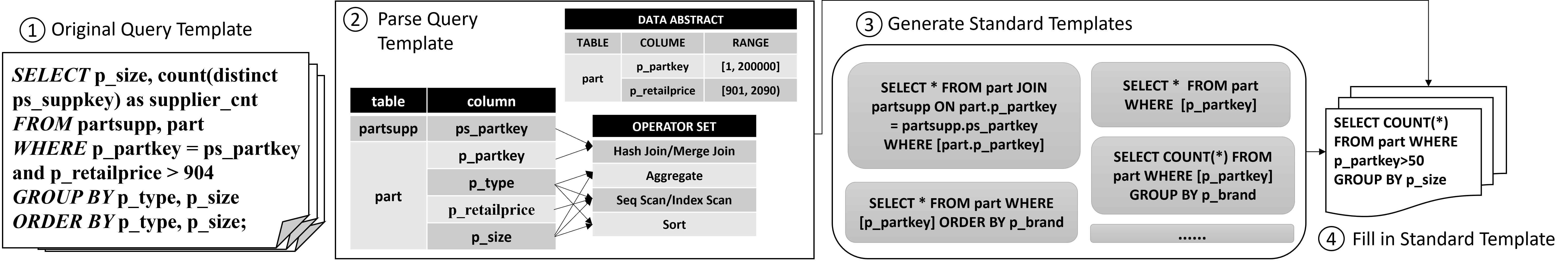}
    \caption{The example for template query generation.}
    \label{fig:temexa}
\end{figure*}

In the first phase (Lines 2-5), we parse the original query templates and obtain the operator-table-column set, defined as $\{o1:[(t1,c1),(t2,c2),...],...\}$, where $o1$ is an operator, $t1,c1$ is one of the related table-column tuples of $o1$. For each query template $p$, we gather its operator set by matching the keywords-operator relationships in Table~\ref{tab:rule}. Specifically, the keyword "$>$" is corresponding to the seq scan operator and the index scan operator. We observe one keyword may related to two physical operators due to the uncertain query optimization. After parsing the original query templates, we obtain the operator-table-column set, which is the basis to generate the simplified templates. For example, Figure~\ref{fig:temexa} shows an example from an original query template to the simplified queries. The original query template consists of some keywords, like $=, >$, order by, group by, etc. Based on these keywords, we obtain the corresponding operator-table-column set represented by some arrows, like partsupp-p\_partkey $\rightarrow$ hash/index scan.

\begin{table}[htbp]	
    \renewcommand{\arraystretch}{1.1}
        \centering 
        \caption{The basic templates of operators.}  
        \label{tab:rule}  
    \begin{tabular}{|m{1.5cm}|m{1.5cm}|m{3cm}|}
        \hline
        Keyword & Operator & Parent Template \\
        \hline
        $>, like,=, <, in, etc.$ & Seq/Index Scan & 
        SELECT * FROM [table] WHERE [condition] \\
        \hline
        Order By & Sort & SELECT * FROM [table] WHERE [condition] ORDER BY [table.attr] \\
        \hline
        Group By & Aggregate & SELECT COUNT(*) FROM [table] WHERE [condition] GROUP BY [attribute] \\
        \hline
        \multirow{2}{1.5cm}[-3ex]{table1.attr1\\=\\table2.attr2} & \multirow{2}{1.5cm}[-4ex]{Merge/Hash Join, Nested loop }  & SELECT * FROM [table1] JOIN [table2] ON [table1.attr = table2.attr] WHERE [condition]\\
        &  & SELECT * FROM [table1] JOIN [table2] ON [table1.attr = table2.attr] WHERE [condition] ORDER BY [table1.attr] \\
        \hline
       ... & ... & ... \\
        \hline
    \end{tabular}
\end{table}

In the second phase (Lines 6-9), we generate the simplified templates according to the parent template correlation in Table~\ref{tab:rule} and the operator-table-column set $info$. For each kind of operator, we design some parent templates to reproduce operators. For the seq/index scan, we design one parent template, ``SELECT * FROM [Table] WHERE [CONDITION]". We do not enforce the operator type in parent templates by database commands, like enable index scan. The reason is that the enforcement may miss the characteristics of the original queries. For example, one column with the corresponding index has a large amount of index scan instead of seq scan. Hence, we generate simplified query templates without some enforced commands. After identifying the parent templates, we generate specific templates by filling in the table-column information. For example in Figure~\ref{fig:temexa}, for the seq scan operator, we generate the template by filling the table-column information into the template "SELECT * FROM partsupp WHERE [ps\_partkey]". 

In the third phase (Lines 11-15), we fill the simplified templates with the data abstract $R$ and random keyword from $\{<, >,=, in, like,...\}$ to iteratively generate the simplified queries. We set up the scale parameter to control the number of template queries. Finally, we obtain the simplified queries, which could be used to calculate the feature snapshot.

In general, the time complexity of this algorithm is related to the length of the $P$, the generated operator length, the length of the table-column pair of operators, and the scale $N$, which are constants. Hence, our simplified query generation algorithm is in constant time complexity. Next, we analyze some limitations of our simplified query templates.

\textbf{Discussions:} In this paper, our feature snapshot is established for the operator level. This is an estimated representation of the ignored variables. Naturally, it could be extended to more fine-grained levels such as the operator-table level and even the operator-table-column level, i.e. we could establish the own feature snapshot for the scan operator of certain tables and certain columns. Fine-grained feature snapshots will bring higher efficiency, and also increase the collection cost. 

%% file: content/selection.tex
Existing AI-driven works~\cite{card3,marcus2019plan} of query cost estimation employ a direct feature encoding for queries and data tables. The large dimension of features brings overhead computational. In this section, we introduce our feature reduction algorithm, a novel method to filter the useless features for AI-driven query cost estimators, to further improve the time-accuracy efficiency. We first analyze the encoding characteristics of existing popular AI-driven database techniques in Section~\ref{sec:encode}. Then, we introduce our difference-propagation feature reduction method in Section~\ref{sec:node}, to filter the useless features. 

\subsection{The analysis of existing encoding methods}\label{sec:encode}
As the foundation of the design feature reduction algorithm, we analyze the encoding characteristics of existing popular AI-driven database techniques. As shown in Table~\ref{tab:plan}, we conclude the coding objects, the coding methods, and the model type for seven popular works varying multiple database tasks. The specific analysis is as follows.

\begin{table}[ht!]	
\renewcommand{\arraystretch}{1.1}
	\centering 
	\caption{The encoding methods of existing works for queries.}  
	\label{tab:plan}  
\begin{tabular}{|m{1.4cm}|m{1cm}|m{1.2cm}|m{0.6cm}|m{1.5cm}|} 
    \hline
	Method & plan or operator & encoding methods & model & task  \\
	\hline 
    QPPnet~\cite{marcus2019plan} &physical operator & one-hot, numerical value & DNN & cost estimation \\
    \hline
    MSCN~\cite{mscn} & query plan & one-hot, numerical value & DNN & cardinality estimaiton \\
    \hline                                                        
    AVGDL~\cite{view} & physical operator & one-hot, numerical value & RNN & view selection \\
    \hline
    end2end~\cite{sun2019end} & physical operator & one-hot, numerical value & RNN & cost estimaiton \& cardinality estimaiton\\
    \hline
    zero-shot~\cite{2022zero} & physical operator & numerical value & MLP &  cost estimaiton  \\
    \hline
    AIMeets~\cite{estimator} & physical operator  & numerical value, one-hot & DNN & Index selection\\
    \hline
    Bao~\cite{marcus2020bao} & physical operator & one-hot, numerical value & tree CNN & query optimization \\
    \hline
    
\end{tabular}
\end{table}

For encoding objects, we find most techniques encoding the query from the aspects of fine-grained operator level, like AIMeets~\cite{estimator}, AVGDL~\cite{view}, etc. Only the MSCN~\cite{mscn} directly encodes the queries from the three aspects, the join, predicates, and data samples. Thus, our feature reduction algorithm should be designed for the fine-grained operator level to adapt to most models. At the same time, the fine-grained design could also be compatible with high-level encoding objects.

For the encoding methods, we could find that all the techniques utilize the same encoding method, the one-hot encoding, and numerical values. For example, the QPPNet~\cite{marcus2019plan} utilizes the plan-structural deep neural network to learn the relationships between the query physical plan and its time cost. For every node of the plan, QPPNet employs the one-hot encoding for node type (like scan, sort, etc.), index, table, and the numerical value for columns, node width, card, and cost. Consequently, our feature reduction algorithm should be compatible with the one-hot encoding and numerical values.

For the model type, the learned models of existing methods consist of three typical networks, deep neural network (DNN), recurrent neural network (RNN), and convolutional neural network (CNN). Then, our feature reduction algorithm should be compatible with different types of neural networks.

Totally, feature reduction needs to consider three core sights, including the fine-grained operator level, the one-hot encoding and numerical values, and the different typical neural networks.

\subsection{Feature Reduction}\label{sec:node}
In this section, we introduce our fine-grained feature reduction algorithm designed for query operators. The feature reduction involves two key factors: the input labeled data and the learned model. The input-labeled data contains the operator-level feature snapshot, the query feature, and the statistics feature. The learned model refers to the AI-driven cost model for query estimation. Due to the different learning abilities, the different models may have different effective features on the same labeled data. The feature reduction problem is an complex problem ($O(2^n)$) due to exponential feature combinations.  The specific problem is defined as follows:
\begin{algorithm}
    \SetKwInOut{Input}{input}
    \Input{
        \ the labeled input data ($D$), the learned cost model ($M$) \\
    }
    \SetKwInOut{Output}{output}
    \Output{
        \ $F$ is the filtered features. \\
    }
    \BlankLine

    $flag \leftarrow 0, C^{min} \leftarrow qerror(M, D)$\\

    \While{$flag == 0$}{
        $drop,C \leftarrow \varnothing, C^{min}$ \\
        \ForEach{$f \in D.X.features$}{
            $C_{f} \leftarrow qerror(M, D.X.reduce(f))$ // \textit{calculate the changed q-error}\\
            \If{$C > C_{f}$}{
                $C \leftarrow C_{f}, drop \leftarrow f $\\
            }
        }
        \If{$Drop == \varnothing$}{
            $flag \leftarrow 1$ \\
        } 
        $D \leftarrow D.X.reduce(Drop)$\\ 
    }
    $F = D.X.features$ \\
    return $F$ \\

\caption{The approximate greedy feature reduction algorithm for query cost estimation}\label{alg:greedy}
\end{algorithm}

\begin{figure}[ht]
    \centering
    \subfigure[An example of query plan encoding.]{
		\includegraphics[width=0.9\linewidth]{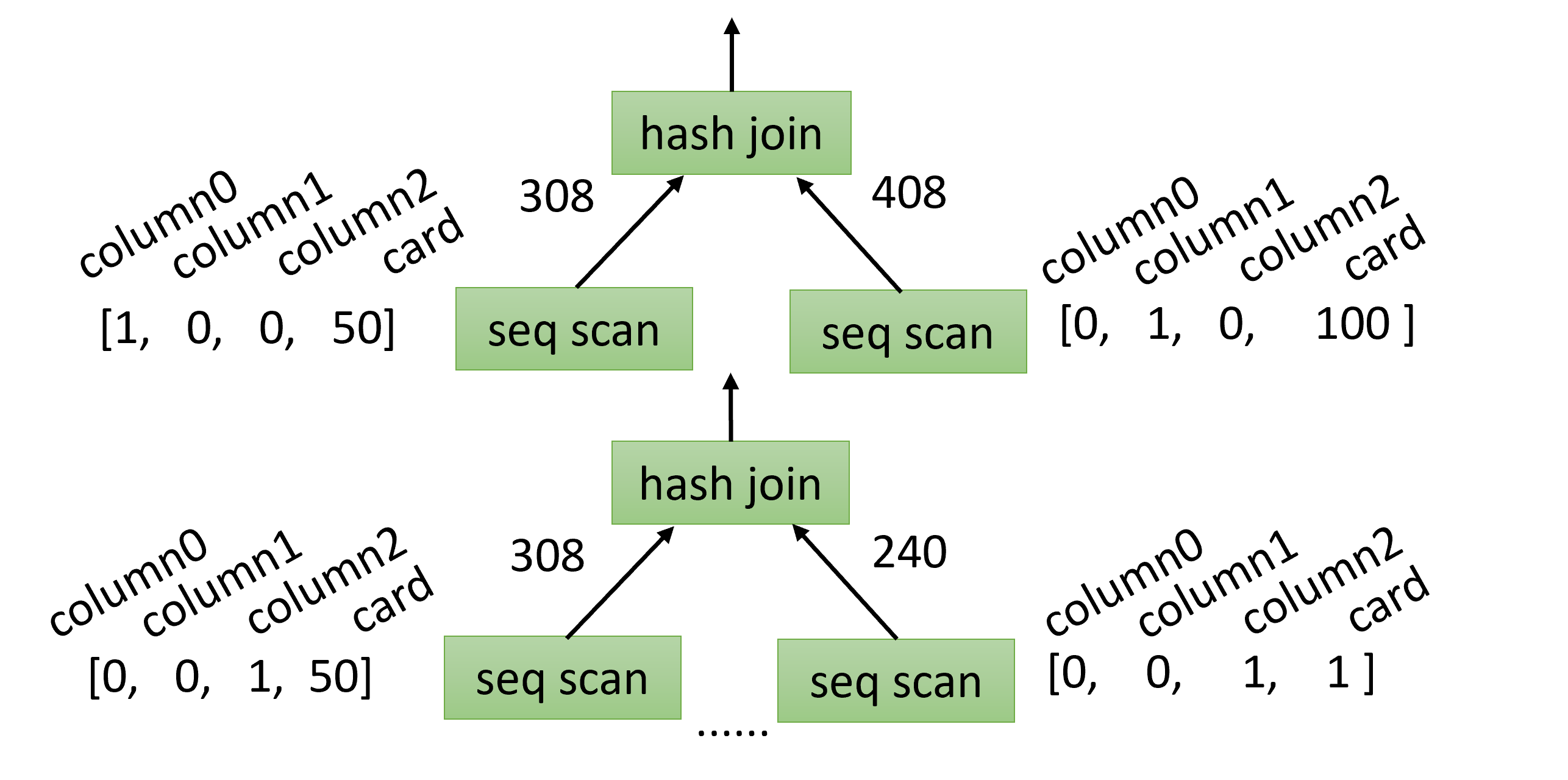}
		\label{fig:exa:plan}
	}
    \subfigure[An example of a learned model.]{
		\includegraphics[width=0.9\linewidth]{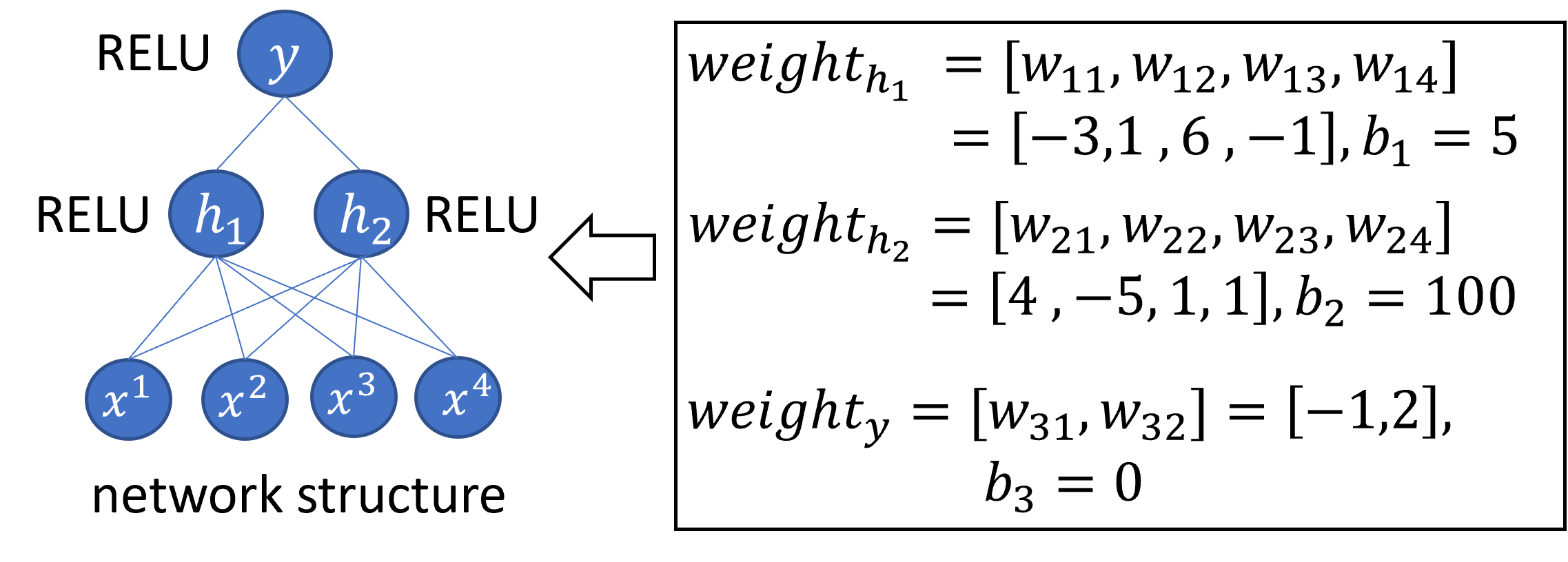}
		\label{fig:exa:model}
	}
    \caption{The basic example for clarifying the feature reduction}
    \label{fig:gradient}
\end{figure}

Given the cost estimation model $M$ and labeled operator set $D = \{X,Y\} = \{(x_1, y_1), .. (x_n, y_n)\}$, where the $x_i$ is the encoding of certain query operator (like scan, sort, etc), $y_i$ is the corresponding cost and $M$ is the corresponding learned model ($Y = M(X)$). The goal of feature reduction is to filter the useful feature input consisting of the one-hot and numerical values. Figure~\ref{fig:gradient} shows a simple example for operator encoding and the learned cost model (the DNN and RELU are also used in existing query cost estimators~\cite{marcus2019plan,sun2019end}). The $D = \{([1,0,0,50],154), ([0,1,0,100],204),([0,0,1,50],150),$\\
$([1,0,0,1],104)\}$ is a labeled data of scan operator. The feature reduction task aims to obtain the effective subset feature of the four dimension input under the three-layer neural network model.

Feature reduction is a classical task of machine learning~\cite{dong2018feature}, which is significant in improving the time-accuracy efficiency of ML models. A typical idea is to greedily select a subset of features from the original feature set to form a new feature set so that the new feature set can be used to train the learned model with better performance. Considering that the combinatorial explosion feature subsets will bring large time complexity, $O(2^n)$. In order to solve this problem within polynomial time, we ignore the co-relationships between features and consider the approximate q-error-based greedy algorithm to find an effective feature subset as shown in Algorithm~\ref{alg:greedy}. In Lines 2-10, we reduce the useless features by iteratively dropping the feature with the minimum q-error. The q-error-based approximate greedy algorithm could reduce the useless features with the polynomial time complexity $O(n^2)$.

Even though the above approximate greedy algorithm could complete feature reduction within polynomial time, ignoring the co-relationships between features may retain some potentially useless features. That is, due to the mutual influence between the two features, removing one feature alone may increase the error, but removing two at the same time will reduce the error.

To ensure the algorithm time complexity and the co-relationship between features, we consider a normal solution to filter the useless features, the gradient. The partial derivative of each dimension could not only reflect the influence of certain dimensions of the input but also combine the co-relationships of features~\cite{gradienti}. Specifically, for all types of AI-driven models (like deep network, convolutional network, etc.), we could utilize the background propagation algorithm to quickly gather the expectation changes (gradient changes) as the influence scores for each dimension. Taking the learned model of Figure~\ref{fig:exa:model} as an example, we could calculate the partial gradient expectation of the on $k_{th}$ dimension ($x^k$) by the equation: $I_{gradient}(k)  = E_{x_i \in D}[\frac{\partial(y)}{\partial(x_i^k)}]  = \frac{1}{len(X)}\sum_{x_i \in X} (\frac{\partial(y)}{\partial(h_1)} * \frac{\partial(h_1)}{\partial(x_i^k)} +\frac{\partial(y)}{\partial(h_2)} * \frac{\partial(h_2)}{\partial(x_i^k)})$

Although the gradient reduction method could represent the effectiveness of certain dimensions of input, two characteristics of existing AI-driven database task models~\cite{view,marcus2020bao, mscn} lead to the gradient reduction ineffective. On the one hand, there exists a large amount of one-hot codes which is a discrete variable. The gradient of discrete variables is ineffective. On the other hand, the activation function suffers gradient vanishing problems such as the RELU in QPPNet~\cite{marcus2019plan}. For example in Figure~\ref{fig:gradient}, $h_1 = relu(x*weight_{h_1}^T + b_1) = max(-3x_1 +x_2 +6x_3 - x_4 +5, 0) $, the gradient of $[1,0,0,50]$ and $[0,1,0,100]$ are zero which is ineffective for feature importance. 

\begin{equation}\label{equ:diff}
    \begin{aligned}
        I_{diff}(k)&= \frac{1}{len(R)*len(D)} *\sum_{x_i \in D, x_j \in R}\\
        &[|\frac{M(x_i)-M(x_j)}{h_1(x_i)-h_1(x_j)} * \frac{h_1(x_i)-h_1(x_j)}{x_i^k-x_j^k} \\
        &+ \frac{M(x_i)-M(x_j)}{h_2(x_i)-h_2(x_j)} * \frac{h_2(x_i)-h_2(x_j)}{x_i^k-x_j^k} |]
    \end{aligned}
\end{equation}

To process the discrete variables and gradient vanishing, we consider the difference-propagation method to replace the gradient propagation, which is also used in other field~\cite{deeplift}. Specifically, we consider sampling some points as the reference basis (defined as $R \subset D$) and utilize the expectation differences between $D$ and $R$ as the importance score. The specific importance score for single layer is defined as $ I_{diff}(k) = E_{x_i \in D, x_j \in R}[|(M(x_i)-M(x_j))/(x_i^k-x_j^k)|]$. For the multiple layer model as shown in Figure~\ref{fig:gradient}, we have the difference-propagation equation as shown in Equation~\ref{equ:diff}.

Also, we observe that the difference-propagation method does not suffer from the discrete variable and the gradient vanishing problem. Taking the Figure~\ref{fig:gradient} as an example, we define the reference set $R = ([1,0,0,1],104)$ the difference-importance of $1th$ dimension of $([0,0,1,50],154)$ is $|\frac{308-208}{0-1} * \frac{0-1}{0-1} + \frac{308-208}{154-104} * \frac{154-104}{0-1}| =200 $, which is effective for feature reduction.

Hence, we utilize the difference-propagation as the core of our feature reduction algorithm as shown in Algorithm~\ref{alg:diffp}. Specifically, Line 1 inits the reference set $R$ by sampling $N$ points from $D$. For each dimension, Line 3 calculates the expectation differences of each dimension of $D$ by the Equation~\ref{equ:diff}. If the expectation difference is larger than zero, Line 4 adds this feature to the filtered feature set. Line 7 returns all the filtered useful features.

\textbf{Time Complexity:} The difference-propagation algorithm could reduce the useless features within the polynomial time complexity $O(|D.X.features||D||R|)$. Among them, $|D.X.features|$ refers to the number of the features. And the $|D||R|$ is the time to calculate the expectation differences, which could also be accelerated by matrix multiplication~\cite{1987matrix}.

\begin{algorithm}
    \LinesNumbered
    \SetKwInOut{Input}{input}
    \Input{
        \ the labeled input data ($D$), the learned cost model ($M$), the number of reference ($N$) \\
    }
    \SetKwInOut{Output}{output}
    \Output{
        \ $F$ is the filtered features. \\
    }
    \BlankLine

    $R \leftarrow D.sample(N), F \leftarrow \varnothing$\\
    \ForEach{$f \in D.X.features$}{
        $score_{f} =  I_{diff}(k)$ \\
        \If{$score_{f}>0$}{
        $F \leftarrow F \cup f$ \\
        }
    }
    
    return $F$ \\

\caption{The feature reduction algorithm based on difference-propagation}\label{alg:diffp}
\end{algorithm}

\textbf{Discussions:}
Our work could flexibly extend to dynamic workloads by designing a recall algorithm according to the inherent value of input features. In fact, besides time-accuracy importance, the input features of the query cost model have some inherent value. That is, although some features are worthless in the current query evaluation task, it may have the potential ability to affect the query cost itself. For example, to improve the time-accuracy efficiency of the query cost model under the write-only workload, our algorithm reduces the useless index feature, which is represented as a one-hot vector. However, with the workload changes (50\% read, 50\% write), the partial index features are effective for estimating the cost of read queries. Then, the partial index features have high potential value with dynamic query cost estimation. Therefore, in dynamic scenarios, researchers and users need not only to consider the time-accuracy efficiency of the cost model but also to consider the inherent value of features to prevent missing important features.

%% file: content/experiment_results.tex
In this section, we conduct extensive experiments for our QCFE. Firstly, in Section~\ref{sec:setting}, we introduce our experimental set-up in detail. Then, we present the comparisons with popular existing works in time-accuracy efficiency with extensive benchmarks in Section~\ref{sec:tsimulate}. We introduce the ablation study to demonstrate the effects of our QCFE with different design choices in Section~\ref{sec:ablation}. We demonstrate the QCFE robustness of parameters in Section~\ref{sec:robust}. Finally, we evaluate the transferability of our feature snapshot in Section~\ref{sec:transfer}.

\subsection{Experimental setup}\label{sec:setting}
In this section, we introduce the foundation of our experiments, including the hardware settings, dataset settings, workload configurations, and evaluation metrics.

 \noindent \underline{\textbf{Hardware settings}} To avoid the competition for resources, we utilize different servers to collect labeled sets of queries and implement training for learned cost models. The specific settings for all the data collection are PostgreSQL 14.4, Intel Core R7 7735HS with 16GB memory, and 512GB hard disk. The model training is executed on Intel Core i7-12700H with 42GB memory, 2.5TB hard disk, and NVIDIA GeForce RTX 3060 Laptop GPU. 

\noindent \underline{\textbf{Experimental Datasets}}
 We utilize three open source benchmarks to evaluate our methods, containing TPC-H~\footnote{https://www.tpc.org/tpch/}, Sysbench~\footnote{https://github.com/akopytov/Sysbench/archive/1.0.zip} and IMDB~\footnote{http://homepages.cwi.nl/~boncz/job/imdb.tgz}, which are widely used in existing works~\cite{sun2019end,mscn}. TPC-H is a practical commercial dataset consisting of eight tables and 22 query templates. In this paper, we employ the scale factor of 1 to implement our comparisons. IMDB is a complex \& large movie review dataset consisting of 21 tables with size = 7GB. Sysbench is a synthetic dataset with a simple structure and we set up the table size = 5000000. 

\noindent \underline{\textbf{Workload Configurations}}
To verify the efficiency of our feature snapshot, we randomly generate 20 database configurations of Postgres 14.4. Then, for each benchmark, we construct the labeled queries (scale = 10,000) under all configurations. We set up different workload settings to unify the amount of labeled queries as follows. For TPCH, we randomly generate $40 \times 22$ queries in each configuration, so there are 17,600 queries in total. For Sysbench, we run the oltp\_read\_only.lua with default time setting = 10s to generate label queries. Then, we obtain $14000 \times 20$ (knob configurations) queries within 200s. For IMDB, we run the $10 \times 70$ queries of job-light~\footnote{https://github.com/andreaskipf/learnedcardinalities} in each configuration and then collect 14,000 labeled queries. Finally, we randomly split the labeled data of every benchmark into 2000, 4000, 6000, 8000, 10000 five data sets to conduct comparative experiments. In all scales, we divide the labeled set into 80\% training set and 20\% test set.

\noindent \underline{\textbf{Metric:}}
In this paper, we utilize extensive metrics to evaluate our methods, including Q-error shown in Equation~\ref{equ:qerror}, the pearson coefficient shown in Equation~\ref{equ:pearson}, the training time, and the inference time. Specifically, the Q-error could clarify the accuracy of query cost estimation. The pearson coefficient could identify the coefficient relationship between the true cost labels and the predicted labels. The training time and inference time could reflect the time efficiency of the learn estimators. 

\begin{equation}\label{equ:qerror}
   qerror(q) = max\{ \frac{actual(q)}{predict(q)}, \frac{predict(q)}{actual(q)}  \}
\end{equation}

\begin{equation}\label{equ:pearson}
    correlation = \frac{Cov(actual, predict)}{\sigma_{actual} * \sigma_{predict}}
\end{equation}

\noindent \underline{\textbf{Implementation:}} We evaluate our feature engineering method on two popular AI-driven estimators, QPPNet~\cite{marcus2019plan} based on the open source implementation~\cite{marcusimp}, and MSCN~\cite{mscn}, which have different kinds of model designs. The QPPNet utilizes a plan-structural DNN for query cost estimation while MSCN designs a single flattened DNN for query cardinality estimation. To extend MSCN to estimate the cost (like End-to-End~\cite{sun2019end,querycost}), we change the output of MSCN from cardinality to query cost and add the fine-grained features (containing the cardinality) same with QPPNet to MSCN. 
Further, we utilize the open-source libraries~\cite{shap} (could be used to calculate the difference-propagation) to implement feature reduction code. The source codes of our experiments are available in github~\footnote{https://github.com/AvatarTwi/query\_cost\_feature\_engineering}.

\subsection{The Effectiveness of our feature engineering}\label{sec:tsimulate} 
In this section, we evaluate the time-accuracy efficiency of our feature engineering under five methods, containing PostgreSQL, QPPNet, MSCN, QCFE(qpp), QCFE(mscn). We integrate our QCFE into the operators of QPPNet (called QCFE(qpp)) and MSCN (called QCFE(mscn)). Due to the different complexity of benchmarks, we set training iteration = 800 for job-light, 400 for TPCH, and 100 for Sysbench. 
Specifically, Table~\ref{tab:box} shows the time-accuracy efficiency on extensive benchmarks with scale = 2000, 4000, 6000, 8000, 10000, including the pearson coefficient, the mean q-error, and the training time of all baselines. 

\begin{table*}[!t]
    \caption{The time-accuracy efficiency of our feature engineering on extensive benchmarks.}
    \label{tab:box}
   
    \resizebox{\textwidth}{!}{
    \renewcommand{\arraystretch}{1.2}
        \begin{tabular}{clclccclccclccclccclccc}
        \toprule
        Dataset &  & Model &  & \multicolumn{3}{c}{2000} &  & \multicolumn{3}{c}{4000} &  & \multicolumn{3}{c}{6000} &  & \multicolumn{3}{c}{8000} &  & \multicolumn{3}{c}{10000}\\
        \cline{5-7} \cline{9-11} \cline{13-15} \cline{17-19} \cline{21-23}
        &  &  &  & \multicolumn{1}{c}{pearson} & \multicolumn{1}{c}{mean}  & \multicolumn{1}{c}{time}  
        &  & \multicolumn{1}{c}{pearson} & \multicolumn{1}{c}{mean}  & \multicolumn{1}{c}{time}
        &  & \multicolumn{1}{c}{pearson} & \multicolumn{1}{c}{mean}  & \multicolumn{1}{c}{time}
        &  & \multicolumn{1}{c}{pearson} & \multicolumn{1}{c}{mean}  & \multicolumn{1}{c}{time} 
        &  & \multicolumn{1}{c}{pearson} & \multicolumn{1}{c}{mean}  & \multicolumn{1}{c}{time}\\  
        \midrule
        \multirow{5}{2em}{TPCH}
        & & \textsf{PGSQL}
        && 0.704 & 819.241 & 1.194	&& 0.741 & 882.319 & 2.394	&& 0.746 & 790.83 & 3.506	&& 0.663 & 1393.472 & 4.734	&& 0.632 & 1179.219 & 5.896\\
        & & \textsf{QCFE(mscn)}
        && \textbf{0.986} & 1.094 & \textbf{8.547}	&& \textbf{0.986} & 1.109 & \textbf{25.635}	&& \textbf{0.998} & 1.106 & \textbf{27.984}  && \textbf{0.997} & \textbf{1.086} & \textbf{31.368} && \textbf{0.997} & 1.11 & \textbf{41.399}\\
        & & \textsf{QCFE(qpp)}
        && 0.985 & \textbf{1.072} & 424.322	&& 0.985 & \textbf{1.089} & 399.598	&& 0.985 & \textbf{1.101} & 391.901	&& 0.979 & 1.108 & 433.513	&& 0.969 & \textbf{1.096} & 417.469\\
        & & \textsf{MSCN}
        && 0.983 & 1.105 & 9.608	&& 0.979 & 1.13 & 27.995	&& 0.988 & 1.126 & 40.615 && 0.988 & 1.105 & 33.314	&& 0.987 & 1.134 & 43.437\\
        & & \textsf{QPPNet}
        && 0.985 & 1.107 & 369.467	&& 0.986 & 1.136 & 345.541	&& 0.984 & 1.111 & 361.47	&& 0.963 & 1.129 & 354.904	&& 0.966 & 1.128 & 365.44\\
        \hline
        \multirow{5}{2em}{Sysbench}
        & & \textsf{PGSQL}
        && 0.224 & 169509.592 & 0.006	&& 0.246 & 175054.294 & 0.011	&& 0.287 & 185715.657 & 0.016	&& 0.269 & 978066.774 & 0.022	&& 0.283 & 938706.491 & 0.027\\
        & & \textsf{QCFE(mscn)}
        && 0.792 & \textbf{1.528} & 8.116	&& \textbf{0.748} & \textbf{1.521} & 15.531	&& \textbf{0.818} & \textbf{1.484} & 32.226	&& 0.776 & \textbf{1.542} & 29.579	&& 0.721 & \textbf{1.57} & 41.045\\
        & & \textsf{QCFE(qpp)}
        && \textbf{0.824} & 1.72 & \textbf{5.52}	&& 0.682 & 1.68 & \textbf{4.965}	&& 0.715 & 2.464 & \textbf{5.329}	&& \textbf{0.857} & 1.868 & \textbf{4.409}	&& \textbf{0.787} & 2.01 & \textbf{4.808}	\\
        & & \textsf{MSCN}
        && 0.698 & 1.734 & 7.183	&& 0.709 & 1.738 & 14.713	&& 0.796 & 1.659 & 21.366	&& 0.606 & 1.804 & 28.616	&& 0.648 & 1.785 & 35.351\\
        & & \textsf{QPPNet}
        && 0.524 & 10.402 & 9.321	&& 0.465 & 8.492 & 9.997	&& 0.488 & 8.891 & 9.463	&& 0.616 & 33.596 & 9.459	&& 0.633 & 32.644 & 10.282\\
        \hline
        \multirow{5}{2em}{job-light}
        & & \textsf{PGSQL}
        && 0.447 & 150.103 & 0.009	&& 0.394 & 171.211 & 0.017	&& 0.396 & 137.072 & 0.033	&& 0.367 & 153.256 & 0.042	&& 0.376 & 148.1 & 0.048\\
        & & \textsf{QCFE(mscn)}
        && 0.985 & \textbf{1.083} & \textbf{7.029}	&& \textbf{0.996} & \textbf{1.066} & \textbf{18.709}	&& 0.996 & \textbf{1.056} & \textbf{25.3}	&& \textbf{0.998} & \textbf{1.046} & \textbf{31.915}	&& \textbf{0.998} & \textbf{1.046} & \textbf{45.077}\\
        & & \textsf{QCFE(qpp)}
        && \textbf{0.994} & 1.18 & 229.067	&& 0.996 & 1.127 & 243.729	&& \textbf{0.997} & 1.162 & 230.444	&& 0.995 & 1.148 & 241.921	&& 0.996 & 1.243 & 255.174\\
        & & \textsf{MSCN}
        && 0.99 & 1.086 & 7.229	&& 0.994 & 1.074 & 33.012	&& 0.993 & 1.074 & 33.012	&& 0.995 & 1.069 & 37.713	&& 0.994 & 1.07 & 68.241\\
        & & \textsf{QPPNet}
        && 0.993 & 1.2013 & 353.156	&& 0.989 & 1.423 & 361.81	&& 0.993 & 1.248 & 333.015	&& 0.985 & 1.445 & 337.923	&& 0.992 & 1.261 & 527.086\\
        \bottomrule
        \end{tabular}
    }
\end{table*}

\textbf{The time-accuracy efficiency of TPCH:} In Table~\ref{tab:box}, compared to MSCN, we observe that our QCFE(mscn) improves the mean q-error from 1.094 to 1.086, the pearson coefficient from 0.985 to 0.993 on average. Compared to PostgreSQL, QCFE(mscn) improves the mean q-error from 1,012.6 to 1.086, the pearson coefficient from 0.697 to 0.993 on average. That clarify that our QCFE could effectively construct useful features for AI-driven query cost model. At the same time, our QCFE(mscn) saves average 12.4\% time efficiency for learned model training. That demonstrates that our QCFE could effectively reduce the useless features and improve model training speed. Especially, on the scale = 10000 for QCFE(mscn), our method improves the q-error from 1.134 to 1.11 and the pearson coefficient from 0.987 to 0.997. Similarly, the QCFE(qpp) also reaches higher time-accuracy efficiency than QPPNet on all scales. 

\textbf{The time-accuracy of job-light:} Compared to TPCH, the job-light has more complex table relationships, and we observe that the average mean q-error is larger than TPCH. Specifically, compared to QPPNet, our QCFE(qpp) improves the mean q-error from 1.316 to 1.172, and the pearson coefficient from 0.990 to 0.996 on average. Compared to PostgreSQL, QCFE(qpp) improves the mean q-error from 151.8 to 1.172, the pearson coefficient from 0.396 to 0.996 on average. That clarifies our QCFE could effectively construct useful features for AI-driven query cost model. At the same time, our QCFE(qpp) saves average 37.3\% time efficiency for learned model training. That demonstrates that our QCFE could effectively reduce the useless features and improve model training speed. Especially, on the scale = 8000 for QCFE(qpp), our method improves the q-error from 1.445 to 1.148 and the pearson coefficient from 0.985 to 0.995. Similarly, the QCFE(mscn) also reaches higher time-accuracy efficiency than MSCN on all scales.

\textbf{The time-accuracy of Sysbench:} Compared to other benchmarks, the Sysbench only has a single table and its query templates have simple structures. Thus, it is more difficult to learn the relationships between query features and query cost. For example, there exists q-error = 10 with the real query time cost = 0.0001 and predict cost = 0.001. We observe that the PostgreSQL, MSCN, and QPPNet have average mean q-error and average pearson. Although Sysbench is difficult to predict, our QCFE(mscn) and QCFE(qpp) still reach pearson coefficient and q-error on average. Also compared to QPPNet, our QCFE(qpp) saves average 16.7\% time efficiency for learned model training. Especially, on the scale = 8000 for QCFE(qpp), our method improves the q-error from 1.804 to 1.542 and the pearson coefficient from 0.616 to 0.857.

\begin{figure}[htb!]
	\centering
		\includegraphics[width=0.95\linewidth]{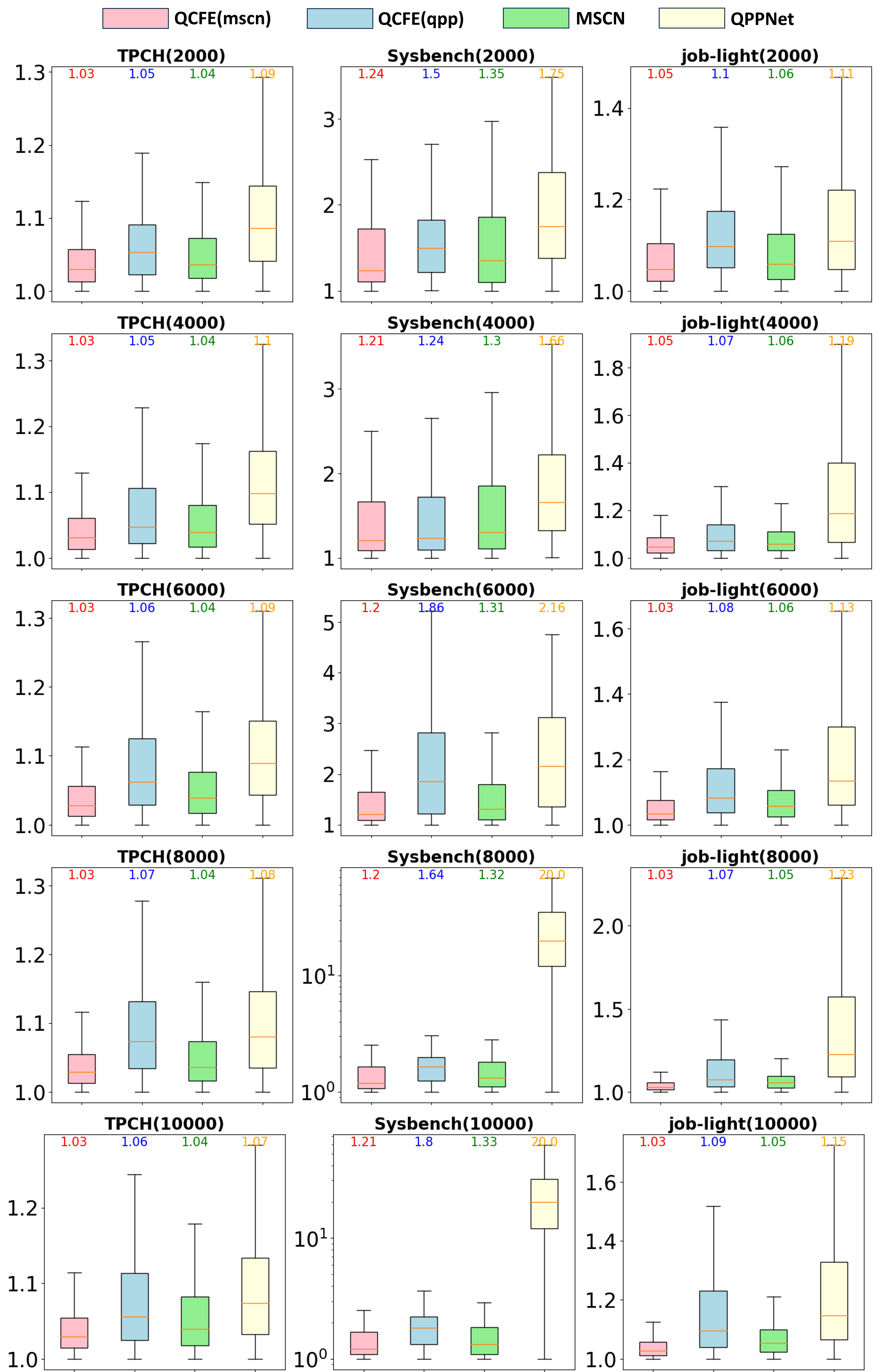}
    \caption{The variance of q-error comparisons. The box boundaries are at the 25th/50th/75th percentiles}
    \label{fig:box1}
\end{figure}

\textbf{The Variance of Q-error:} Moreover, we present the q-error variance of different benchmarks on scale = 2000, 4000,  6000, 8000, 10000 in Figure~\ref{fig:box1}. Totally, Our method can effectively reduce the variance of the q-error in extensive benchmarks, making the prediction effect more stable and accurate. Compared to QPPNet, we observe that our QCFE(qpp) reduces the q-error to 57.143\% (from 1.084 to 1.048), 0.032\% (from 9.16 to 1.308), 50.299\% (from 1.167 to 1.084) at the 50\% quantiles under TPCH, Sysbench, job-light. For example, in the TPCH with scale = 2000, our QCFE(qpp) can reduce the 90\% percentiles q-error from 1.3 to 1.2 compared to QPPNet. Also, other benchmarks have similar comparison results.  Furthermore, we have observed that as the amount of data increases, the QPPNet and MSCN both show deterioration in prediction effects and increase in error variance, while our method can ensure higher accuracy and better variance under various data amounts. Hence, our method could effectively construct useful features for learned estimators and reduce the variance of q-error.

\subsection{Ablation Study}\label{sec:ablation}
In this experiment, we present the results of our ablation study to show the effects of our QCFE under the different design choices, including FSO (the feature snapshot calculated by original queries), FST (the feature snapshot from template queries), FSO + FR, FSO + GD(gradient) and FSO + Greedy. We focus on the effectiveness of the different feature snapshots and various feature reduction methods. 
\begin{figure}[htb]
    \centering
    \includegraphics[width=0.9\linewidth]{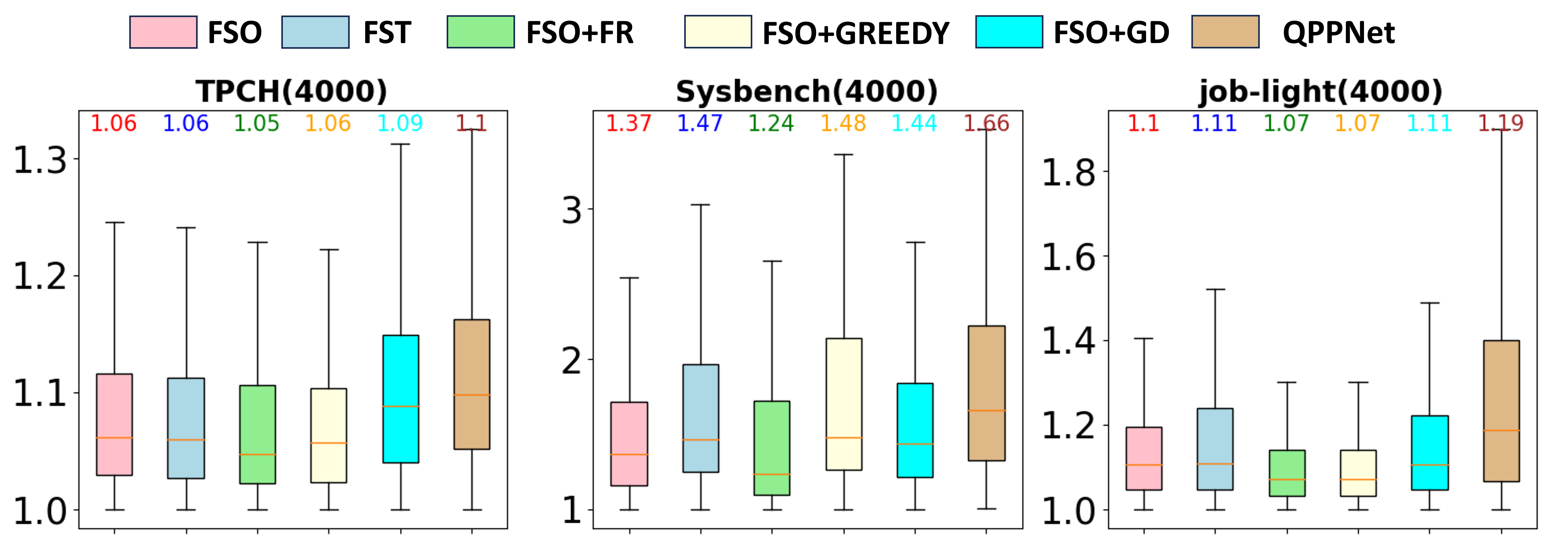}
    \caption{The ablation study of our QCFE on extensive benchmarks with scale = 4000.}
    \label{fig:ablation}
\end{figure}

Figure~\ref{fig:ablation} shows the results of our ablation study of QPPNet model on extensive benchmarks with scale = 4000. We observe that the FST could reach the similar [1.109,1.781,1.222] mean q-error with FSO [1.098,1.715,1.180]. This demonstrates that our standard templates could reflect the characteristics of original queries. Moreover, we observe that our difference-importance outperforms Greedy and GD in both TPCH and job-light. This is because FR could effectively reduce the useless features and improve the accuracy of learned estimators. While GD may suffer the discrete one-hot codes and the gradient vanishing. The greedy method Greedy only reduces one dimension in one iteration. Thus, Greedy may miss the co-relationships between features.

\begin{figure*}[htb]
    \centering
    \includegraphics[width=0.95\linewidth]{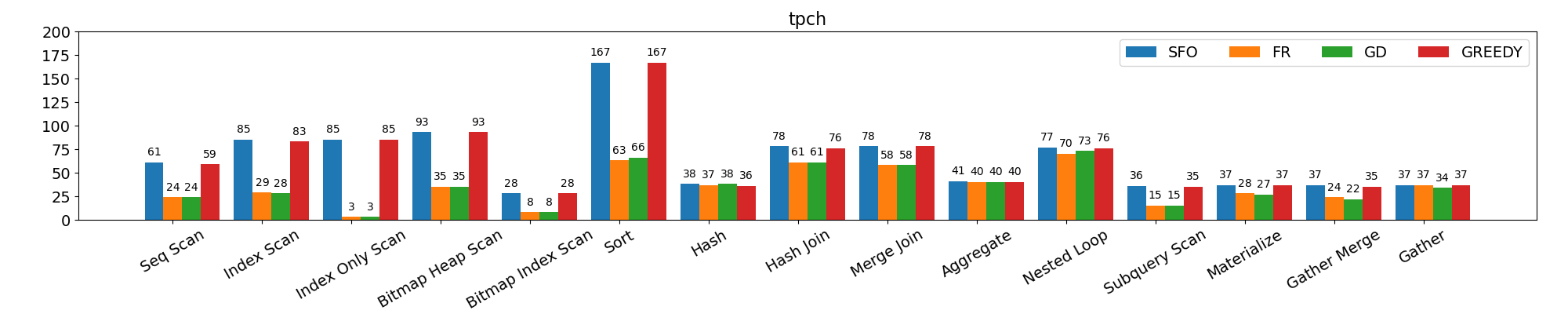}
    \caption{The ablation study of our QCFE on extensive benchmarks with scale = 4000.}
    \label{fig:feature}
\end{figure*}
Moreover, we present the feature reduction results for TPCH in Figure~\ref{fig:feature}. We observe that Greedy reduces above 1.20\% features, GD above 41.22\%, and FR above 41.22\% on average. Among them, the Greedy retains the majority of features because it cannot capture the co-relationships between features, like the pair of useless features. Specifically, our FR reduces up to 57 features in index scan operator while Greedy only reduces 2 features. This demonstrates the feature of index scan exists multiple. Moreover, we find that GD also could reduce a large number of features. In particular, GD could reduce the 101 features of the sort operator. This is because the GD could capture the co-relationships between features. However, due to the one-hot codes and gradient vanishing, its reduction may produce wrong importance scores. The 50th q-error of GD in Figure~\ref{fig:ablation} only reaches 1.44 while FR could reach 1.24.

\subsection{The robustness of parameters}\label{sec:robust}

In this section, we demonstrate the robustness of two important parameters of QCFE, the scale of templates and the number of reference.

\begin{table}[htb]
    \caption{}
	\tabcolsep=0.05cm
    \label{tab:robust1}
    \scalebox{0.9}{
	\begin{tabular}{clcclcclcclcclcc}
        \toprule
        TYPE &  & \multicolumn{2}{c}{FSO} &  &\multicolumn{2}{c}{1(2)} &  & \multicolumn{2}{c}{2(4)} &  & \multicolumn{2}{c}{3(6)}&  & \multicolumn{2}{c}{4(8)}\\ 
        \cline{3-4} \cline{6-7} \cline{9-10} \cline{12-13} \cline{15-16}
        &  & \multicolumn{1}{c}{mean} & \multicolumn{1}{c}{time}   
        &  & \multicolumn{1}{c}{mean} & \multicolumn{1}{c}{time}    
        &  & \multicolumn{1}{c}{mean} & \multicolumn{1}{c}{time}   
        &  & \multicolumn{1}{c}{mean} & \multicolumn{1}{c}{time}   
        &  & \multicolumn{1}{c}{mean} & \multicolumn{1}{c}{time}   \\
        \midrule
        \textsf{TPCH}
        && 1.098 & 7.7h 
        && 1.106 & \textbf{0.8h} 
        && 1.110 & 1.9h 
        && 1.108 & 2.3h
        && \textbf{1.096} & 3.8h \\
        \textsf{job-light}
        && \textbf{1.18} & 31.8h 
        && 1.262 & \textbf{0.9h}
        && 1.222 & 1.7h 
        && 1.210 & 2.6h 
        && 1.187 & 3.5h
        \\
        \bottomrule
    \end{tabular}
    }
\end{table}

As shown in Table~\ref{tab:robust1}, we observe that the effects of different scales of templates. Totally, we conclude that our simplified SQL templates could reflect the characteristics of original queries. And the q-error is relatively robust with the changes of scales. For TPCH, our FS generates 123 SQL templates, and we set scale = 1,2,3,4. We observe that calculating the feature snapshot costs 7.7h for labeling queries with original templates of TPCH. While our simplified templates only cost 3.8h and reach a competitive q-error (1.096) with the FSO. For job-light, FS generates 19 SQL templates due to its light join structure. Hence, we set up the scale = 2,4,6,8 to generate more queries. Because the multiple join in job-light has large time consumption, our simplified templates will reduce the queries collect time to 3.5h/31.8h $\approx$ 11\% and also reach a competitive q-error.

\begin{table}[h!]
	\centering
	\caption{The effects of the number of reference in TPCH (scale = 2000) and QCFE(qpp).}
	
	\tabcolsep=0.1cm
	\label{tab:robust2}
		\begin{tabular}{cccccccc}
		\toprule
		number  & mean & q-error95 &q-error90 & runtime/s & reduction ratio\\
		\midrule
		\textsf{200}
		 & 1.107  & 1.39 & 1.224 & 267.788 & 40.036\%\\
		\textsf{250}
		& 1.09  & 1.305 & 1.212 & 267.788 & 39.857\%\\
		\textsf{300}
		 & 1.095  & 1.262 & 1.212 & 349.73 & 40.214\%\\
		\textsf{400}
		 & 1.09  & 1.27 & 1.196 & 591.671 & 40.125\%\\
		\textsf{500}
		  & 1.076  & 1.215 & 1.156 & 911.671 & 39.767\%\\
		\bottomrule
	\end{tabular}
\end{table}

Table~\ref{tab:robust2} shows the robustness of the number of references in TPCH and QCFE(qpp), including the mean q-error, the 95th q-error, the 90th q-error, the runtime of FR, and the reduction ratio. We observe that the q-error is slightly improved with the reference number increasing. This clarifies that the limited reference could reflect the importance of input features. And the runtime increases linearly with parameter changes, from 267s to 911s. This demonstrates that our QCFE has a linear increase with the changes of reference. Further, the reduction ratio (about 40\%) is also robust with the change of reference number. 

\subsection{The transferability of feature snapshot}\label{sec:transfer}

In this section, we introduce the transferability of our feature snapshot. Specifically, we utilize the trained cost model (scale = 10000 in TPCH and job-light) of the above hardware settings as the transferable basis. Then, in the same TPCH and job-light settings, we collect 2000 labeled queries as the training set and 500 labeled queries as the test set in a new hardware setting (Intel Core i7-12700H with 42GB memory, 2.5TB hard disk), called h2. To clarify the transferability of our feature snapshot, we calculate the feature snapshot with the 2000 labeled original queries (FSO) and the template queries (FST scale = 4 in TPCH and 4 in job-light). 

Table~\ref{tab:trans} shows pearson coefficient, the mean q-error, and the training time of the cost model directly trained with 2000 labeled data, the basis model + FSO with 200 retraining, and the basis model + FST with 200 retraining. We find that only by replacing the feature snapshot and a little retraining our transferable model could reach a similar accuracy to the model trained with 2000 labeled data. Especially, we observe that the FST outperforms the FSO in mean q-error under TPCH. Because TPCH has more various operators (123 templates). It is difficult for 2000 labeled queries to calculate a representative feature snapshot. And our FST could cover more operator characteristics than the FSO with limited labeled set. 

\begin{table}[h!]
\renewcommand{\arraystretch}{1.1}
	\centering
		\caption{The transferability of feature snapshot in TPCH and job-light.}
		
		\label{tab:trans}
			\begin{tabular}{clclclc}
			\toprule
			Model &  & Metric &  & TPCH &  & job-light\\
			\midrule
			\multirow{3}{*}{basis}
			&& \textsf{pearson}
			&& 0.983 && 0.995 \\
			&& \textsf{mean}
			&& 1.088 && 1.195 \\
			&& \textsf{time}
			&& 381.157 && 232.519 \\
			\hline
			\multirow{3}{*}{trans-FSO}
			&& \textsf{pearson}
			&& 0.981 && 0.997 \\
			&& \textsf{mean}
			&& 1.112 && 1.246 \\
			&& \textsf{time}
			&& 114.455 && 65.539 \\
			\hline
			\multirow{3}{*}{trans-FST}
			&& \textsf{pearson}
			&& 0.982 && 0.99 \\
			&& \textsf{mean}
			&& 1.083 && 1.278 \\
			&& \textsf{time}
			&& 121.093 && 73.246 \\
			\bottomrule
		\end{tabular}
	\end{table}

\begin{figure}[htb]
    \centering
    \includegraphics[width=0.9\linewidth]{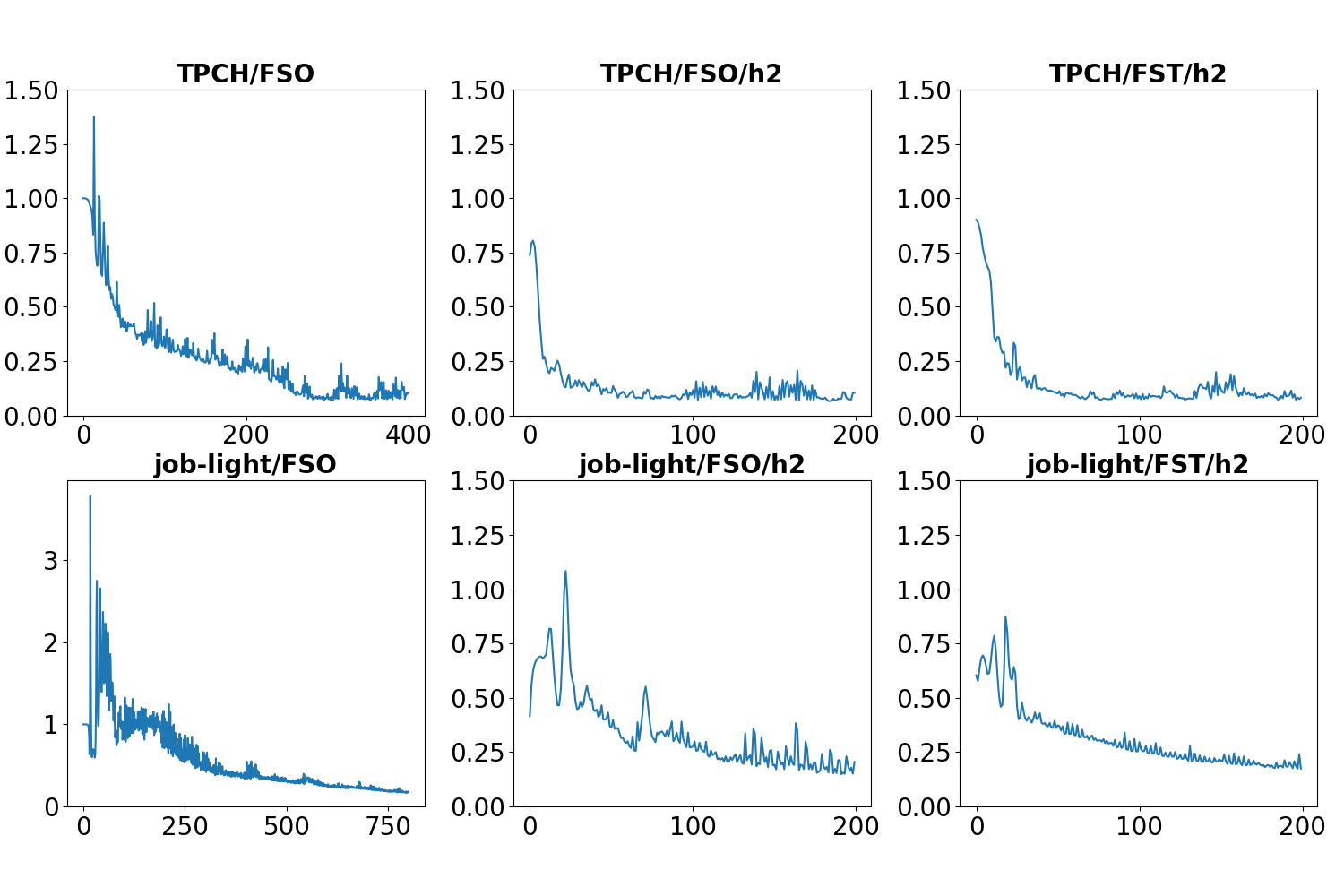}
    \caption{The prediction error changes of the transferable learned models with the iteration.}
    \label{fig:know:dif}
\end{figure}

Moreover, we demonstrate the convergence speed of the direct training model and the transferable model in Figure~\ref{fig:know:dif}. We observe that the transferable model could reach a similar accuracy with the direct training model with 25\% training time. This demonstrates that our feature snapshot could achieve hardware transferability and effectively reduce the training time with a new database environment.

%% file: content/related.tex
In this section, we introduce the related works from two aspects: the cost estimation methods and the feature processing methods. 

\textbf{Cost Estimation:} In the early days, in order to improve the efficiency of query optimization, researchers proposed some statistical methods~\cite{2012Robust, 2014Uncertainty, wu2013predicting} to hypothetically estimate the query performance. For example, Li et al.~\cite{2012Robust} proposed operator-based statistical techniques to estimate query execution time. This model simulates operator costs by modeling resource usage as operator-level behavior and designing a statistical learning model. Modeling load performance as a random variable, Wu et al.~\cite{2014Uncertainty} proposed a model that provides uncertain information for query execution time prediction. The model expresses the cost of a query as a function of the selectivity of operators in the query plan and factors that describe system CPU and I/O operation costs. Recently, the database community has attempted to utilize deep learning model~\cite{2022conditional, yu2022cost} to improve the accuracy of simulating the query performance. For example, Sun et al.~\cite{card3} proposed an end-to-end framework for learning cost estimation, TPool, which uses a tree-structured model to approximate query costs. This research proposes effective feature extraction and encoding techniques to improve the learning ability of the model. Marcus et al.~\cite{marcus2019plan} proposed a deep neural network QPPNet based on the query plan tree structure to estimate the query cost. The model can automatically discover the characteristics of operators and query plans, avoiding manual feature engineering. Akdere et al~\cite{akdere2012learning} proposed predictive modeling techniques to learn query execution behavior at different granularities, from coarse-grained plan-level models to fine-grained operator-level models. This method can well balance the two indicators of accuracy and transferability. Moreover, the zeroshot~\cite{2022zero} focused on the query encoding issues, and proposed a transferable feature across databases as the basis of zero-shot estimation.

\textbf{Feature Engineering of cost estimation:} Feature engineering is a typical task for improving the efficiency of deep learning. There exist many feature engineering methods~\cite{nargesian2017learning,scott1999feature} in other fields, like the work2vec~\cite{word2vec} in natural language processing. As for query cost estimation, there only exist two relative works, the transferable query encoding~\cite{2022zero} and query representation learning~\cite{queryformer}. (1) The zeroshot~\cite{2022zero} focused on the poor migration of query encoding and converted the original one-hot encoding of attributes into the encoding of attribute types. This encoding method supports a transferable feature across databases as the basis of zero-shot estimation. (2) The QueryFormer~\cite{queryformer} proposed a tree-transformer model to learn the tree structure of queries. The goal of QueryFormer is to translate query execution plans into vectorized representations. Totally different from the above two works, our QCFE aims to find the effective features for query cost estimation, which can be used to improve the time-accuracy efficiency of query cost estimation. And our method can be easily combined with the above two methods and enhance their efficiency.

%% file: content/conclusion.tex
In this paper, we propose QCFE to obtain effective features for query cost estimation. On the one hand, we design an estimated feature snapshot to integrate some other ignored variables, containing database knobs, hardware, etc. The experimental comparisons clarify that our approach could efficiently capture the influence of ignored variables. On the other hand, we design a difference-propagation feature reduction method to filter the useless features, to improve the time-accuracy efficiency of query cost estimation. 
In the future, we consider optimizing QCFE from the following two aspects. On the one hand, we consider adapting our feature reduction algorithm to the dynamic workload situations by designing a recall mechanism. On the other hand, we aim to enhance the transfer learning of query cost estimation by identifying which feature to transfer.